\documentclass[3p,times,twocolumn]{elsarticle}

\usepackage{ecrc}
\def\d{\mathrm{d}}
\usepackage{slashed}
\usepackage{amsmath}
\usepackage{subfigure}
\def\d{\mathrm{d}}
\def\beq{\begin{equation}}
\def\eeq{\end{equation}}
\def\bea{\begin{eqnarray}}
\def\eea{\end{eqnarray}}

\def\nnb{\nonumber}

\def\nnb{\nonumber}

\def\ba{\begin{array}}
	\def\ea{\end{array}} 
\volume{00}

\firstpage{1}

\journalname{Nuclear and Particle Physics Proceedings}

\runauth{}

\jid{nppp}

\jnltitlelogo{Nuclear and Particle Physics Proceedings}

\usepackage{amssymb}

\usepackage[figuresright]{rotating}

\begin{document}

\begin{frontmatter}



\dochead{}

\title{Application of AdS/QCD to B physics}

\cortext[cor0]{Talk given at 18th International Conference in Quantum Chromodynamics (QCD 15,  30th anniversary),  29 june - 3 july 2015, Montpellier - FR}
\author[label1]{Mohammad Ahmady\fnref{fn1}}
\fntext[fn1]{Speaker, Corresponding author.}
\ead{mahmady@mta.ca}
\address[label1]{Department of Physics, Mount Allison University, Sackville, New Brunswick, Canada E4L 1E6}
\author[label2]{S\'{e}bastien Lord\corref{cor1}}
\cortext[cor1]{Senior undergraduate student.}
\ead{esl8420@umoncton.ca}
\address[label2]{D\'epartement de Math\'ematiques et Statistique, Universit\'e de Moncton,
	Moncton, New Brunswick, Canada E1A 3E9}
\author[label1,label3]{Ruben Sandapen}
\ead{rsandapen@mta.ca}
\address[label3]{Department of Physics, Acadia University, Wolfville, Nova-Scotia, Canada, B4P 2R6}

\begin{abstract}
We build on the success of AdS/QCD in predicting
diffractive $\rho$ meson production to obtain the transition
form factors for B decays to light mesons. Consequently, we predict several observables associated with $B\to \rho\ell\nu$ semileptonic and $B\to K^*\mu^+\mu^-$ dileptonic decays.  Our predictions are compared to the available experimental data.
\end{abstract}


\end{frontmatter}


\section{Introduction}
Rare B meson decays are excellent venues for precision test of the standard model (SM) and search for new physics (NP) beyond it. The underlying quark processes for these decays are suppressed either due to small Cabbibo-Kobayashi-Mashawa (CKM) quark mixing or lack of flavor changing neutral current (FCNC) at tree level in SM.  For example, the semileptonic $B\to\rho\ell\nu$ decay can help with our knowledge of the least known CKM matrix element $V_{ub}$.  On the other hand, a rare B decay like $B\to K^*\mu^+\mu^-$ occurs only through quantum loops, and is sensitive to new unknown particles that are predicted by various NP scenarios beyond the SM.  These so-called exotic particles can appear as virtual entries in the loops and alter the decay rate and other associated observables from what is expected within the SM.  This indirect search for NP in rare B meson decays has been used to constrain various scenarios like supersymmetry and vector-like quark model.

In this talk, we report on a number of predictions for semileptonic $B\to\rho\ell\nu$ and dileptonic $B\to K^*\mu^+\mu^-$ decays based on the anti-de Sitter/Quantum Chromodynamics (AdS/QCD)\cite{Brodsky, rev}.  The holographic light-front wavefunctions obtained from AdS/QCD are used to calculate the distribution amplitudes (DAs) of the $\rho$ and $K^*$ vector mesons\cite{PRD1,PRD2}.  These DAs are then inserted in light-cone sum rules (LCSR) formulas for $B\to \rho ,\; K^*$ transition form factors\cite{PRD3,PRD4,PRD5}.

\section{Transition form factors}

The nonperturbative QCD effects appear in the $B\to V,\; V=\rho ,\; K^*$ transition matrix elements which are parametrized in terms of a number of form factors.  Understanding these form factors is crucial in theoretical calculation of the exclusive B meson decays to light mesons.  Some or all of the 7 form factors defined below appear in semileptonic or dileptonic B decays: 
\begin{figure}
	\centering
	\subfigure[~Twist-$2$ DA for the longitudinally polarized $\rho$ meson]{\includegraphics[width=.30\textwidth]{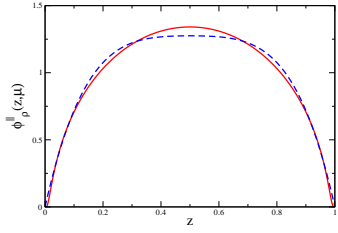} }
	\subfigure[~Twist-$2$ DA for the transversely polarized $\rho$ meson]{\includegraphics[width=.30\textwidth]{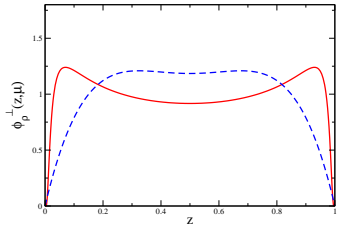} }
	\caption{Twist-$2$ DAs for the $\rho$ meson. Solid: AdS/QCD DAs; Dashed: Sum Rules DAs.} \label{fig:tw2DAs}
\end{figure}
\begin{figure}
	\centering
	\subfigure[~Twist-$2$ DA for the longitudinally polarized $K^*$ meson]{\includegraphics[width=.35\textwidth]{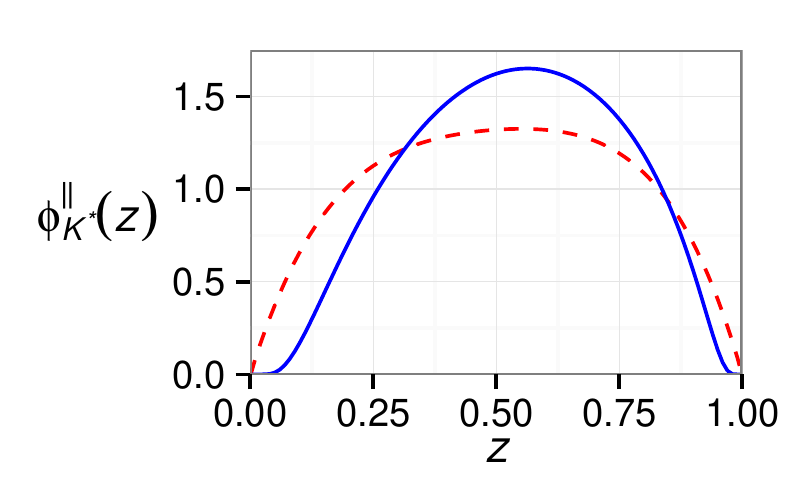} }
	\subfigure[~Twist-$2$ DA for the transversely polarized $K^*$ meson]{\includegraphics[width=.35\textwidth]{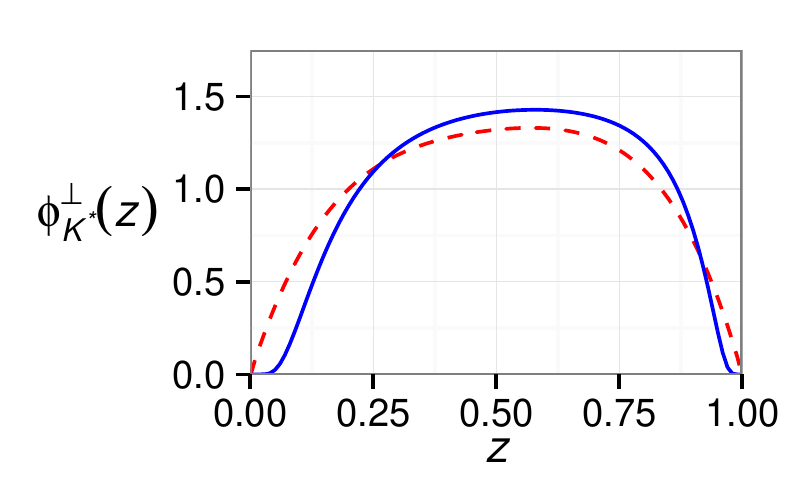} }
	\caption{Twist-$2$ DAs for the $K^*$ meson. Solid: AdS/QCD DAs; Dashed: Sum Rules DAs.} \label{fig:tw2DAsKstar}
\end{figure}
\begin{align}
&\langle V (k,\varepsilon)|\bar{q} \gamma^\mu(1-\gamma^5 )b | B(p) \rangle\nnb \\ &= \frac{2i V(q^2)}{m_B + m_{K^*}} \epsilon^{\mu \nu \rho \sigma} \varepsilon^*_{\nu} k_{\rho} p_{\sigma} -2m_{K^*} A_0(q^2) \frac{\varepsilon^* \cdot q}{q^2} q^{\mu}  \nonumber \\
&- (m_B + m_{K^*}) A_1(q^2) \left(\varepsilon^{\mu *}- \frac{\varepsilon^* \cdot q q^{\mu}}{q^2} \right) \nonumber \\
&+ A_2(q^2) \frac{\varepsilon^* \cdot q}{m_B + m_{K^*}}  \left[ (p+k)^{\mu} - \frac{m_B^2 - m_{K^*}^2}{q^2} q^{\mu} \right] 
\nonumber
\end{align}

\begin{align}
&q_{\nu} \langle V (k,\varepsilon)|\bar{q} \sigma^{\mu \nu} (1-\gamma^5 )b | B(p) \rangle\nnb \\ &= 2 T_1(q^2) \epsilon^{\mu \nu \rho \sigma} \varepsilon^*_{\nu} p_{\rho} k_{\sigma} \nonumber \\
&- i T_2(q^2)[(\varepsilon^* \cdot q)(p+k)_{\mu}-\varepsilon_{\mu}^*(m_B^2-m_{K^*}^2)] \nonumber \\
&- iT_3(q^2) (\varepsilon^* \cdot q) \left[ \frac{q^2}{m_B^2-m_{K^*}^2} (p+k)_{\mu} -q_{\mu}  \right] 
\label{formfactors}
\end{align}
We use LCSR\cite{ali,lcsr} to calculate the seven form factors in terms of the DAs of the light vector mesons. LCSR are variations of the traditional QCD Sum Rules whereby non-local matrix elements are expanded in terms of light front DAs.  Using light-cone coordinates, $x^{\pm}=x^0\pm x^3$, $x^\perp =x^1,\; x^2$, the two twist-2 DAs $\phi_{\perp , \parallel}$ along with the two twist-3 DAs $g_\perp ^{(v,a)}$ are defined through the following relations\cite{Ball:2007zt}:
\begin{align}
&\langle 0|\bar q(0)  \gamma^\mu q(x^-)| V
(P,\lambda)\rangle \nonumber \\
&= f_{V} M_{V}
\frac{e_{\lambda} \cdot x}{P^+x^-}\, P^\mu
\int_0^1 \mathrm{d} u \; e^{-iu P^+x^-}
\phi_\parallel(u,\mu)\nnb
\\ 
&+ f_{V} M_{V}
\left(e_{\lambda}^\mu-P^\mu\frac{e_{\lambda} \cdot
	x}{P^+x^-}\right)
\int_0^1 \mathrm{d} u \; e^{-iu P^+x^-} g_\perp^{ (v)}(u,\mu)\nnb \;, \\
&\langle 0|\bar q(0) [\gamma^\mu,\gamma^\nu] q (x^-)|V
(P,\lambda)\rangle \nnb \\ 
&=2 f_{V}^{\perp} (e^{\mu}_{\lambda} P^{\nu} -
e^{\nu}_{\lambda} P^{\mu}) \int_0^1 \mathrm{d} u \; e^{-iuP^+ x^-} \phi_{\perp}
(u, \mu)\nnb  \;, \\
&\langle 0|\bar q(0) \gamma^\mu \gamma^5 s(x^-)|V (P,\lambda)\rangle \nnb \\
&=-\frac{1}{4} \epsilon^{\mu}_{\nu\rho\sigma} e_{\lambda}^{\nu}
P^{\rho} x^{\sigma}  \tilde{f}_{V} M_{V} \int_0^1 \mathrm{d} u \; e^{-iuP^+ x^-}
g_\perp {(a)}(u, \mu)  \;,\nnb \\
\label{DA}
\end{align}
where 
\begin{equation}
\tilde{f}_{\rho} = f_{\rho}\;,
\end{equation}
and
\begin{equation}
\tilde{f}_{K^*} = f_{K^*}-f_{K^*}^{\perp} \left(\frac{m_s + m_{\bar{q}}}{M_{K^*}} \right) \;.
\end{equation}	
$m_s$ and $m_{\bar q}$ are the masses of the strange quark and light anti-quark, respectively.  Also, all the above-defined DAs are normalized, i.e.
$$
\int_0^1 \d u \;
\phi_V^{\perp ,\parallel}(u,\mu)=	\int_0^1 \d u \; g^{\perp (a,v)}_V(u, \mu)=1
$$
As $x^-\to 0$, in Eqn. \ref{DA} we recover the usual definition for the decay constant $f_V$ and $f^\perp_V$:
\begin{equation}
\langle 0|\bar q(0)  \gamma^\mu q(0)|V
(P,\epsilon)\rangle
= f_V M_V \epsilon^\mu \; ,
\label{decayconstant}
\end{equation}
\begin{equation}
\langle 0|\bar q(0) [\gamma^\mu,\gamma^\nu] q(0)|V (P,\epsilon)\rangle =2 f_V^{\perp} (\epsilon^{\mu} P^{\nu} - \epsilon^{\nu} P^{\mu})\; .
\label{tensordecayconstant}
\end{equation}

The DAs which are commonly used in the literature are derived from QCD sum rules (SR).  In the next section, we put forward alternative DAs which are obtained from AdS/QCD.

\section{AdS/QCD holographic DAs}
The holographic light-front wavefunction for a vector meson $(L=0,S=1)$ in AdS/QCD can be written as\cite{Brodsky,Brodsky2}:
\begin{eqnarray}
\phi_{\lambda} (z,\zeta) &=&{\mathcal N}_\lambda \sqrt{z(1-z)} \exp
\left(-\frac{\kappa^2 \zeta^2}{2}\right)
\nnb \\
&\times&\exp\left \{-\left[\frac{m_q^2-z(m_q^2-m^2_{\bar{q}})}{2\kappa^2 z (1-z)} \right] 
\right \} \label{AdS-QCD-wfn}
\end{eqnarray}
with $\kappa=M_{V}/\sqrt{2}$ and $m_q=m_{\bar q}$ or $m_s$ for $\rho$ or $K^*$ respectively.  $\lambda$ is the polarization of the vector meson.  We have introduced the dependence on quark masses following a prescription by Brodsky and de T\'eramond \cite{Brodsky2}.  This wavefunction for the $\rho$ meson was successfully used to predict diffractive $\rho$ meson electroproduction at HERA\cite{PRL}.  

The relation between the wavefunction (\ref{AdS-QCD-wfn}) and DAs for $\rho$ and $K^*$ was derived in \cite{PRD1,PRD2}.  Figures \ref{fig:tw2DAs} and \ref{fig:tw2DAsKstar} compare the two twist-2 DAs for these two vector mesons obtained from AdS/QCD and SR.

\begin{figure}
	\centering
	{\includegraphics[width=.30\textwidth]{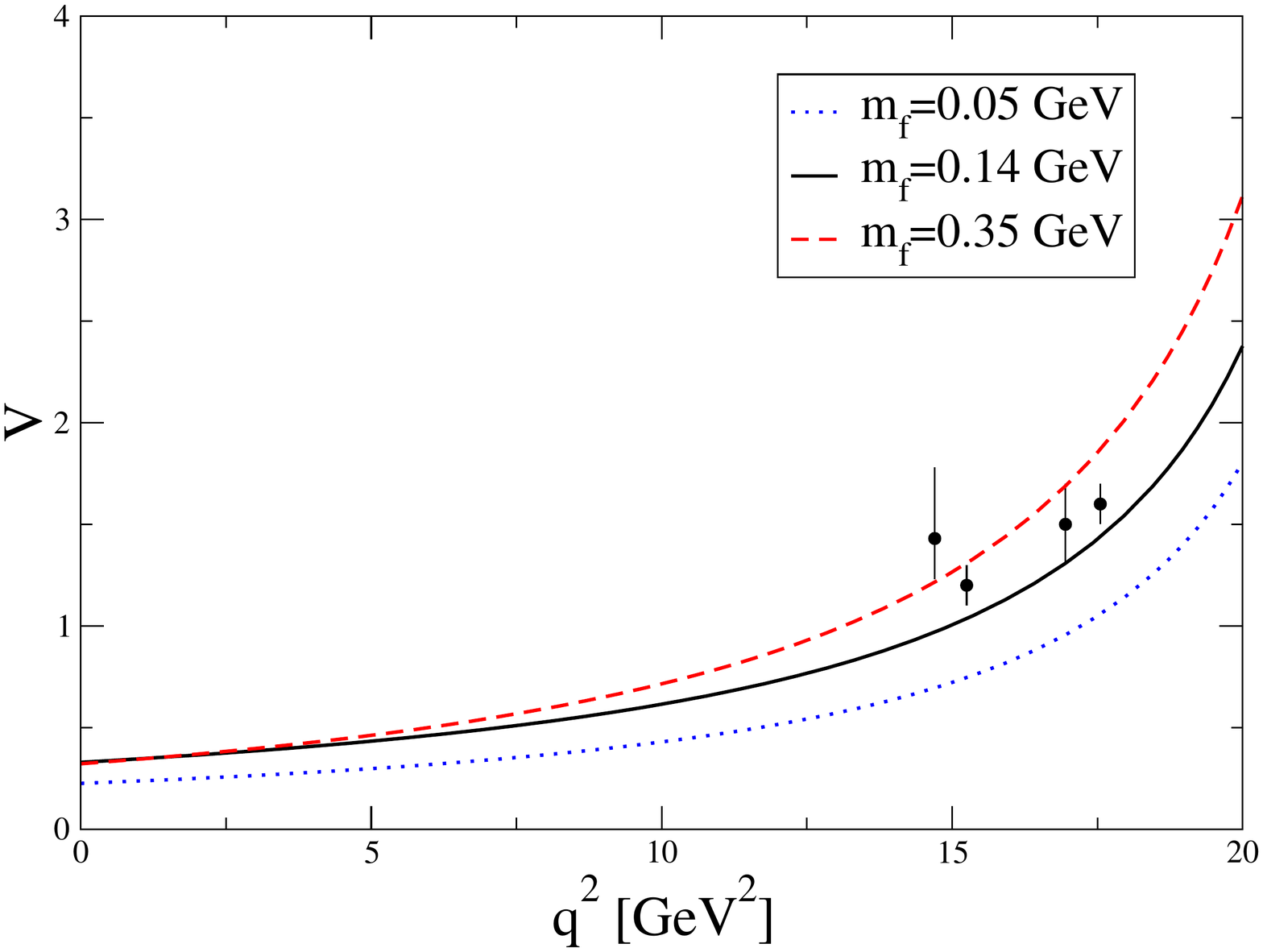} }
	{\includegraphics[width=.30\textwidth]{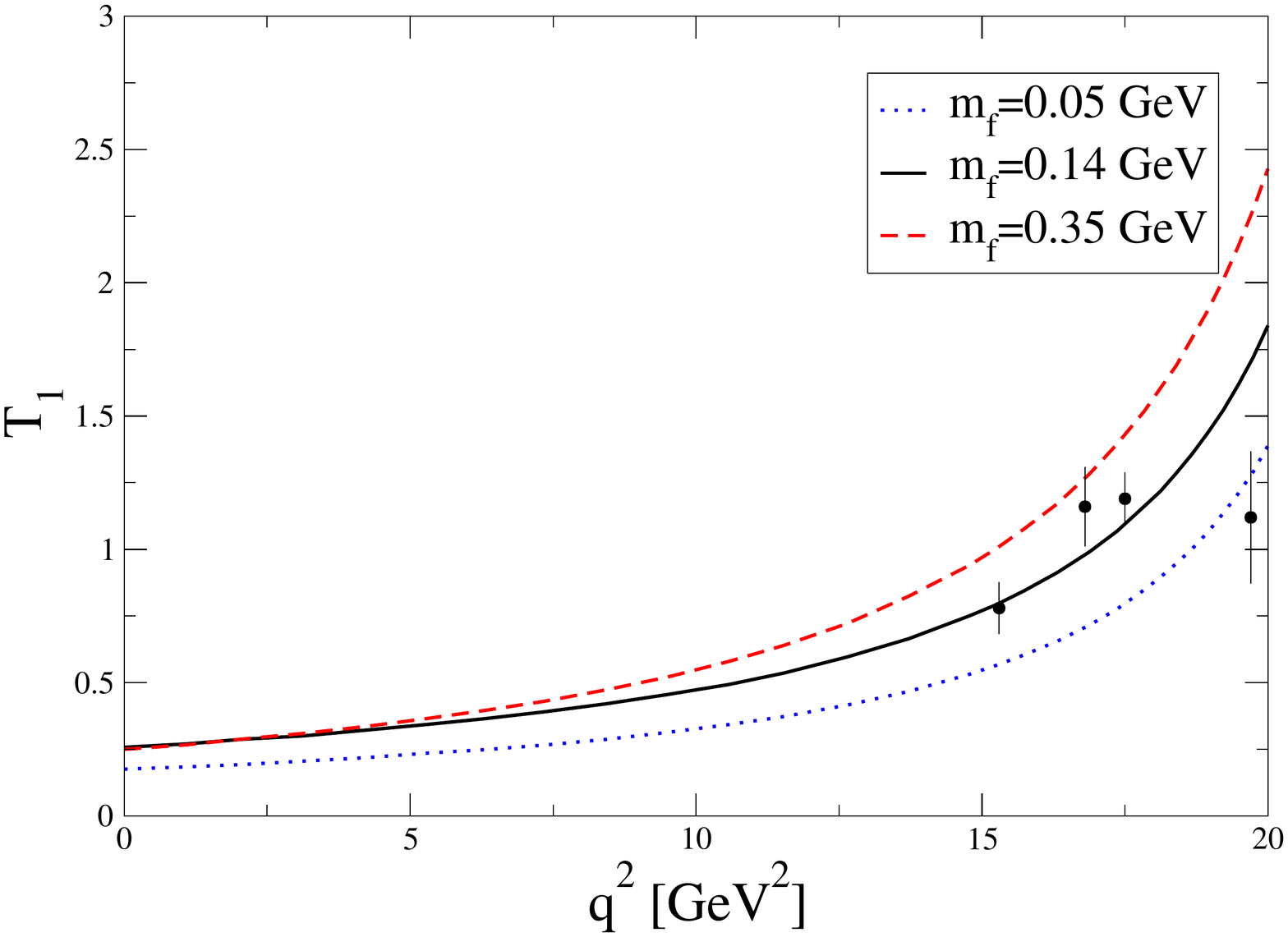} }
	\caption{The AdS/QCD prediction for $B\to\rho$ transition form factors $V$ and $T_1$ for 3 different quark mass inputs.  The available lattice data at high $q^2$ are shown as well.  } \label{formfactorsrho}
\end{figure}

Consequently, the AdS/QCD predictions for $B\to\rho ,\; K^*$ form factors can be computed via LCSR.  One should note that LCSR results are valid at low to intermediate values of the momentum transfer $q^2$.  Figure \ref{formfactorsrho} illustrates the AdS/QCD predictions for two $B\to\rho$ form factors $V$ and $T_1$ when 3 different values of the quark mass are used.  Our results for the all 7 form factors for this transition can be found in \cite{PRD3}.  The data points on this figure are from lattice calculation which are available at high $q^2$\cite{Lattice}.  To calculate the differential branching of the semileptonic $B\to\rho\ell\nu$ decay, we find two-parameter fits for the form factors using AdS/QCD predictions at low to intermediate $q^2$ and lattice data at high $q^2$\cite{PRD3}. 

\begin{figure}
	\centering
	{\includegraphics[width=.30\textwidth]{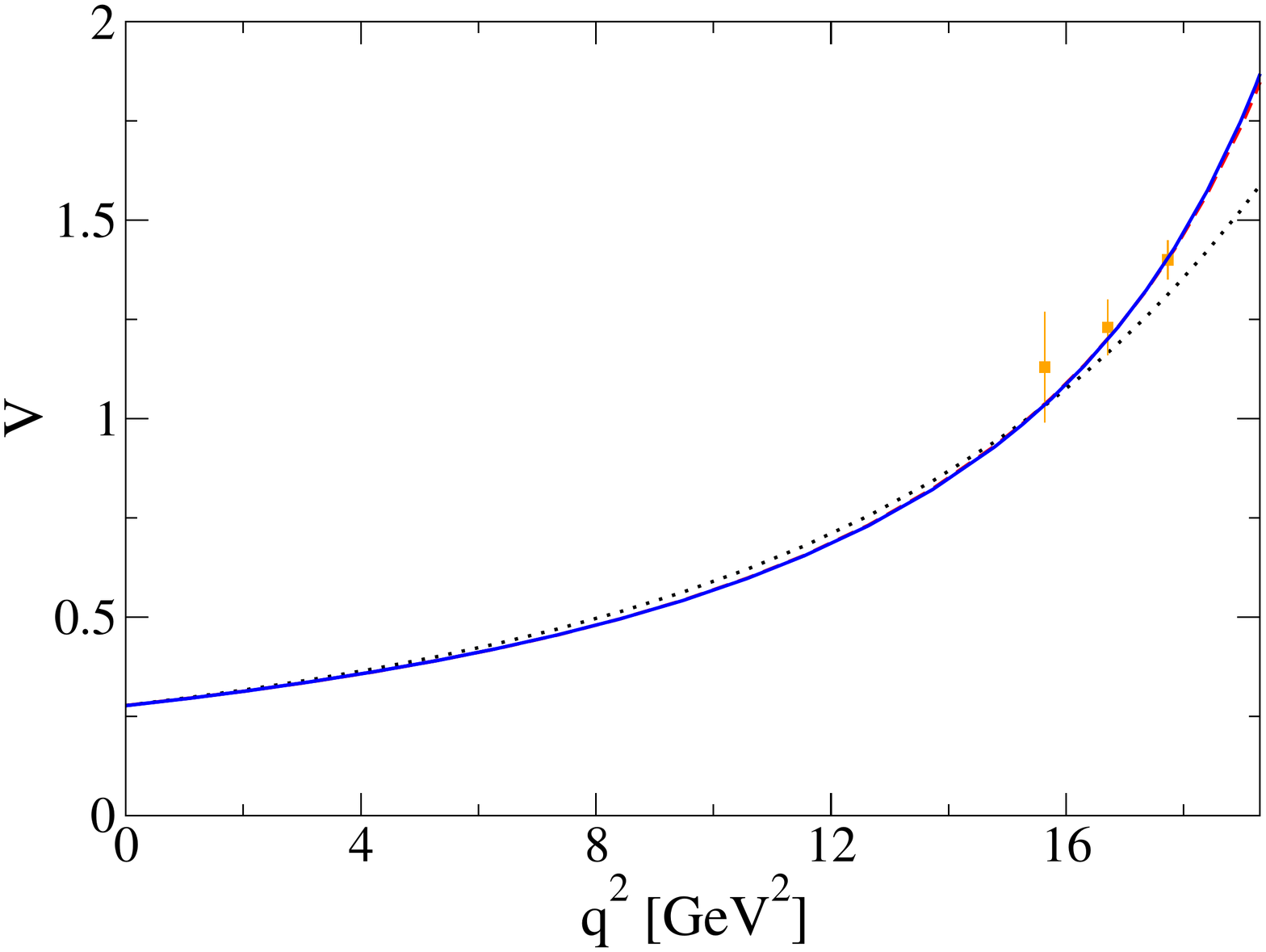} }
	{\includegraphics[width=.30\textwidth]{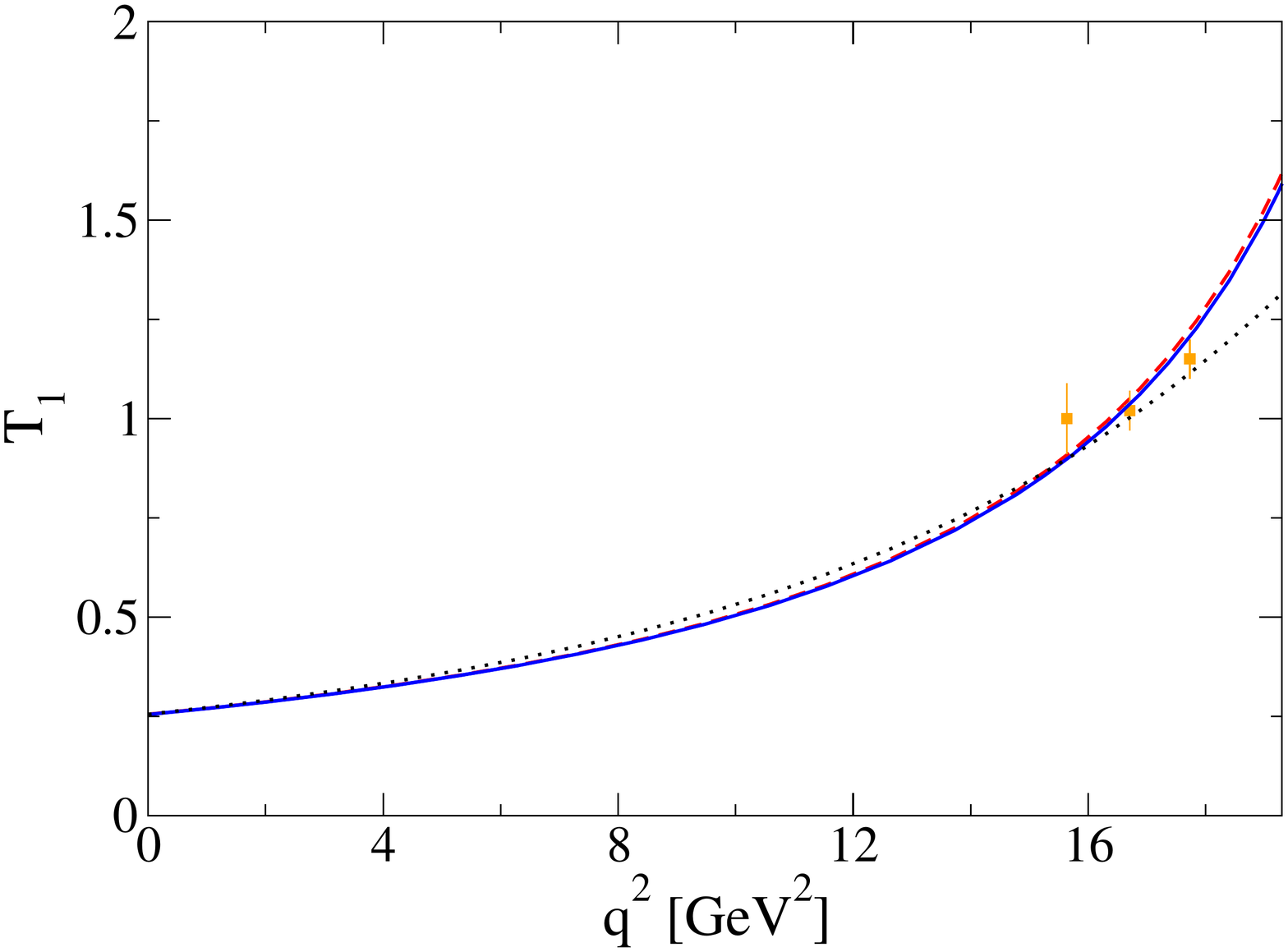} }
	\caption{The AdS/QCD  prediction for $B\to K^*$ transition form factors $V$ and $T_1$. The solid curve denotes AdS/QCD. The dashed curve denotes the AdS/QCD fit. The dotted curve denotes the fit to AdS/QCD and lattice. The available lattice data at high $q^2$ are shown as well. } \label{formfactorsKstar}
\end{figure}

Figure \ref{formfactorsKstar} shows the AdS/QCD prediction for $B\to K^*$ form factors $V$ and $T_1$ as compared with those obtained from SR.  The data points at high $q^2$ are from lattice calculations\cite{latticekstar}.  Our results for the full set of $B\to K^*$ transition form factors can be found in \cite{PRD4}.  Similar to $B\to\rho$ case, we use two-parameter fits to AdS/QCD and lattice data combined for the form factors and use them in our numerical computations.

\section{Numerical predictions for observables}

The differential decay rate of the semileptonic $B\to \rho\ell\nu$ is sensitive to $V_{ub}$.  Figure \ref{fig:tw2DAs} shows our AdS/QCD prediction for the differential decay rate divided by $|V_{ub}|^2$ for two different quark masses.  The data points on this figure are lattice predictions generated by UKQCD Collaboration\cite{Lattice}.

BaBar collaboration has measured the partial branching fractions for this decay channel in three $q^2$ bins\cite{babar}: $\Delta B_{\mbox{\tiny{low}}}=(0.564\pm 0.166)\times 10^{-4},\; \Delta B_{\mbox{\tiny{mid}}}=(0.912\pm 0.147)\times 10^{-4}$ and $\Delta B_{\mbox{\tiny{high}}}=(0.268\pm 0.062)\times 10^{-4}$ for $0<q^2<8,\; 8<q^2<16$ and $16<q^2<20.3\; {\rm GeV}^2$ respectively.  To eliminate the uncertainty in $V_{ub}$ when comparing our results with the above data, we take the ratios of the partial branching fractions as defined below:
\begin{eqnarray}
R_{\mbox{\tiny{low}}}&=&\frac{\Delta B_{\mbox{\tiny{low}}}}{\Delta B_{\mbox{\tiny{mid}}}}=0.618\pm 0.207 \; ,\nnb \\ R_{\mbox{\tiny{high}}}&=&\frac{\Delta B_{\mbox{\tiny{high}}}}{\Delta B_{\mbox{\tiny{mid}}}}=0.294 \pm 0.083\; .
\end{eqnarray}
Our AdS/QCD predictions for these ratios are $R_{\mbox{\tiny{low}}}=0.580, 0.424$ and $R_{\mbox{\tiny{high}}}=0.427,0.503$ for $m_q=0.14$ GeV, $0.35$ GeV.  It seems that better agreement with data is achieved at low $q^2$.

\begin{figure}
	\centering
	{\includegraphics[width=.30\textwidth]{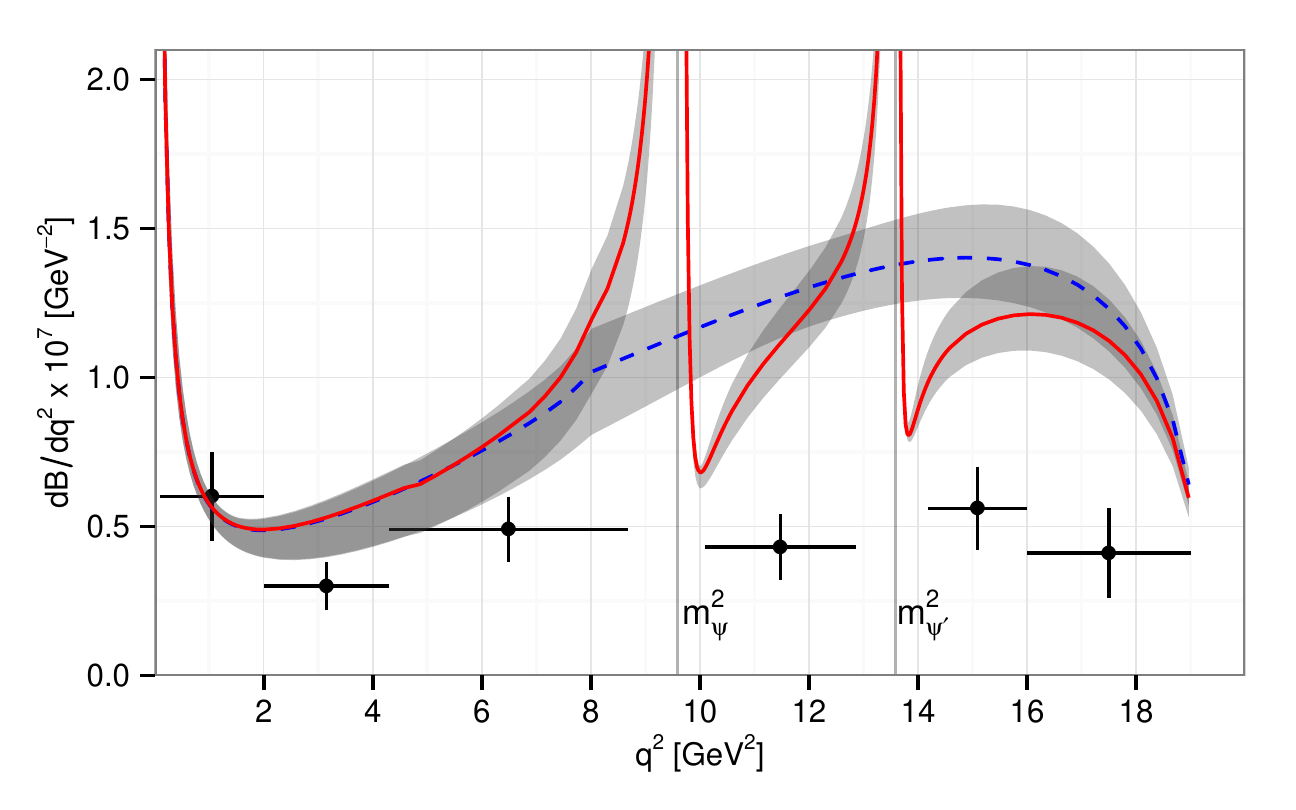}}
	{\includegraphics[width=.30\textwidth]{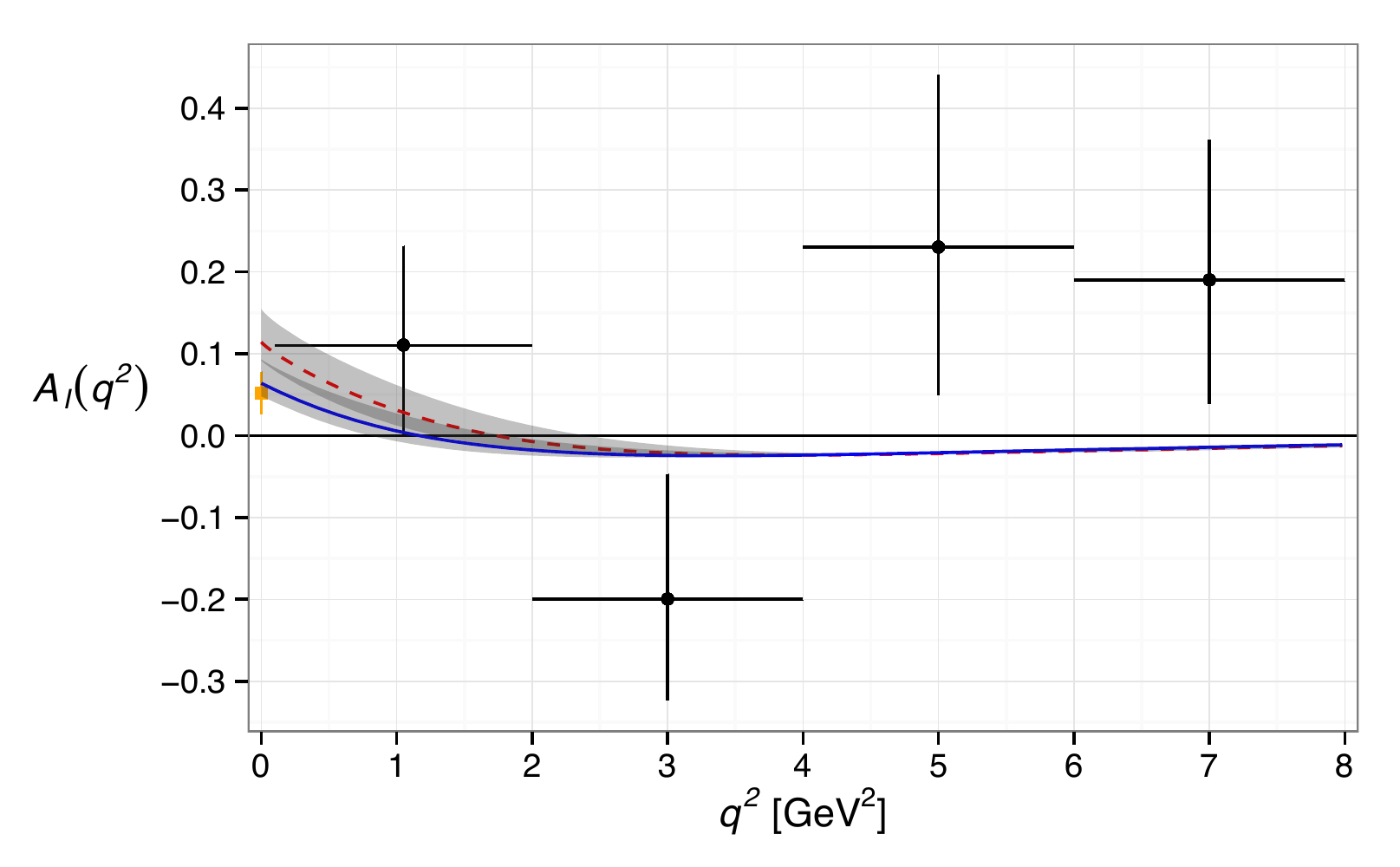} }
	\caption{The AdS/QCD  prediction for $B\to K^*\mu^+\mu^-$ differential branching ratio and isospin asymmetry distribution.  The differential branching ratio is predicted with (solid curve) and without (dashed curve) $c\bar c$ resonances.  In the isospin asymmetry graph, AdS/QCD prediction (solid curve) is presented with an uncertainty band and compared with SR expectation (dashed curve). } \label{BKstarmu}
\end{figure}

Our results for the differential branching ratio and isospin asymmetry distribution in $B\to K^*\mu^+\mu^-$ are shown in Figure \ref{BKstarmu}.  Data points for the branching ratio are from LHCb\cite{LHCb} and the prediction is given with (solid curve) and without (dashed curve) the $c\bar c$ resonance contributions.  Our AdS/QCD prediction for the isospin asymmetry distribution in $B\to K^*\mu^+\mu^-$ in Figure \ref{BKstarmu} is presented (solid curve) with an uncertainty band due to the renormalization scale variation from $m_b/2$ to $2m_b$. The data points are from LHCb and PDG\cite{LHCb2,PDG}.  For comparison, we also include the SR predictions (dashed curve) for isospin asymmetry distribution.  Our prediction for asymmetry at $q^2=0$ (relevant to $B\to K^*\gamma$) is consistent with experimental data.

\begin{figure}
	\centering
	{\includegraphics[width=.30\textwidth]{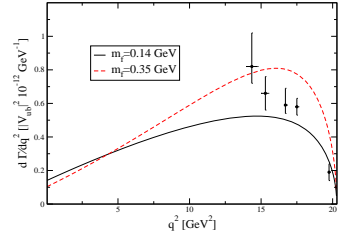} } 
	\caption{Differential decay rate for the semileptonic $B\to\rho\ell\bar\nu$ decay. The lattice data points are from UKQCD Collaboration.}
	\label{fig:tw2DAs}
\end{figure}  

\section{Acknowledgements} %

This research is supported by a team grant from the Natural Sciences and Engineering Research Council of Canada (NSERC).  SL thanks the government of New-Brunswick for SEED-COOP funding.







\end{document}